\RequirePackage[2024-02-02]{latexrelease}
\documentclass[aps,prc,twocolumn,floatfix,superscriptaddress]{revtex4}
\usepackage{epsfig}
\usepackage{amsmath}
\usepackage{color}
\usepackage{graphicx} 
\usepackage{multirow}
\usepackage{booktabs}
\usepackage{subcaption}

\usepackage{caption}
\usepackage{ragged2e}
\begin{document}

\title{Deep Learning for Longitudinal Asymmetry in Heavy-Ion Collisions at RHIC Energies}

\author{Anup Kumar Sikdar}
\email{akssikdar@gmail.com}

\affiliation{Variable Energy Cyclotron Centre, 1/AF, Bidhan Nagar, Kolkata 700064, India}

\author{Sanchari Thakur}
\email{sanchari.thakur@vecc.gov.in}

\affiliation{Variable Energy Cyclotron Centre, 1/AF, Bidhan Nagar, Kolkata 700064, India}

\author{Pingal Dasgupta}
\email{pdasgupta@ifae.es}

\affiliation{Institut de Fisica d’Altes Energies (IFAE) - The Barcelona Institute of Science and Technology, Campus UAB, Bellaterra (Barcelona) Spain}

\author{Rupa Chatterjee}
\email{rupa@vecc.gov.in}

\affiliation{Variable Energy Cyclotron Centre, 1/AF, Bidhan Nagar, Kolkata 700064, India}
\affiliation{Homi Bhabha National Institute, Training School Complex, Anushaktinagar, Mumbai 400094, India}

\begin{abstract}
 We developed a deep neural network framework to predict the longitudinal asymmetry and the corresponding rapidity shift in relativistic heavy-ion collisions. The model was trained using sufficiently large datasets for Au+Au and Cu+Cu collision systems at center-of-mass energies of 62.4 and 200 AGeV at RHIC.  It successfully learns the underlying relationships between the input parameters and reconstructs the corresponding target quantities with good precision. Its predictions agree well with theoretical simulations across both trained and untrained centrality intervals. The influence of longitudinal asymmetry on final-state flow observables has also been studied, and the model results are in reasonable agreement with the simulation. 
 A SHAP-based correlation analysis further supports the physical consistency of the learned feature dependencies and suggests that the observed correlations are not primarily driven by feature redundancy.
The proposed framework provides a computationally efficient approach to estimating event-by-event results, potentially reducing the computational cost of large-scale theoretical model simulations. 
This work presents an initial study exploring the use of deep neural networks to investigate longitudinal asymmetry and rapidity shifts in relativistic heavy-ion collisions.

\end{abstract}

\maketitle

\section{Introduction}

Relativistic heavy-ion collisions provide a unique experimental means of creating and studying the Quark-Gluon Plasma (QGP), a deconfined state of strongly interacting matter believed to have existed during the early stages of the Universe, a few microseconds after the Big Bang ~\cite{qgp1,qgp2,qgp3,qgp4,qgp5}. The initial conditions and the properties of the hot and dense QGP are inferred from a variety of final-state observables. Relativistic hydrodynamics and microscopic transport models provide event-by-event descriptions of heavy-ion collisions and are widely used to interpret experimental data from RHIC and LHC energies ~\cite{Gale,Schenke,Tribedy,Huovinen,ampt,tr1,tr2}. However, a systematic exploration of theoretical results across collision energies, centrality intervals, and system sizes requires a large number of high-statistics calculations. Additionally, the computational demand often limits such studies to a relatively coarse grid of parameter points, making it challenging to explore smooth or non-monotonic trends in the final state observables.

Machine learning surrogate models can help overcome this limitation. Once trained on a representative set of events, a Deep Neural Network (DNN) can predict observables efficiently without running new simulations.
The DNN frameworks have become valuable tools  for learning complex mappings between model parameters and physical observables in various branches of physics. They have been applied to a wide range of problems, including particle identification, detector response, event classification, as well as the prediction of physical observables. In relativistic heavy ion physics, DNNs have been used to analyze collision events, infer properties of the initial state, and predict observables that are otherwise computationally expensive to obtain. Their ability to learn from large datasets makes them particularly useful for studying the complex and fluctuating dynamics of heavy ion collisions.

To interpret the physics learned by a DNN, it is important to examine how the input features influence its predictions. Correlations and SHAP (SHapley Additive exPlanations)   analysis help quantify these dependencies and assess whether the learned relationships are consistent with the underlying physical processes~\cite{lundberg2017}.

In this paper, we present a deep neural network trained on datasets from a microscopic transport model AMPT (A Multi Phase Transport) simulations to study the longitudinal asymmetry in relativistic heavy ion collisions. The model uses the rapidity, charged particle multiplicity, collision energy, and collision system as input features.
It is trained over two collision energies, $\sqrt{s_{NN}}$=62.4 and 200 GeV at RHIC, for both Au+Au and Cu+Cu systems, across three representative centrality classes (0--5\%, 30--40\%, and 50--60\%) spanning central, mid-central, and peripheral collision geometries with 50K events for each configuration which results in a combined training sample of approximately 24 million data points.

The paper is organized as follows. Section II describes the observables and the AMPT model, while Section III presents the network architecture and training procedure. Section IV evaluates the performance of the model and its validation, including correlation analysis and SHAP-based interpretability studies. Section V discusses the physics implications of the results. Finally, Section VI summarizes the main findings and provides an outlook for future applications, including model to data statistical inference.

\section{Longitudinal asymmetry using AMPT model}

In collisions of identical nuclei, the numbers of participants from the two nuclei can fluctuate from event to event due to fluctuations in the positions of individual nucleons around their average nuclear density distributions. This difference in the number of participating nucleons leads to a non-zero net longitudinal momentum of the nucleon–nucleon center-of-mass system, resulting in a rapidity shift of the participant zone relative to the nucleon–nucleon center-of-mass rapidity ~\cite{vov,acharya,raniwala1,raniwala2,Thakur}. This is known as longitudinal asymmetry and is characterized by the magnitude of rapidity shift $y_{0}$. If A and B are the number of participants from two nuclei, the longitudinal asymmetry in participants is defined as,

\begin{equation}
    \alpha_{part} = \frac{(A - B)}{(A + B)}
\end{equation}

and the corresponding rapidity shift is given by ~\cite{acharya,raniwala1},
\begin{equation}
    y_{0} \approx \frac{1}{2} ln(\frac{A}{B})  = \frac{1}{2} ln(\frac{1+\alpha_{part}}{1-\alpha_{part}})
\end{equation}

For small $\alpha_{part}$, $y_{0} \approx \alpha_{part}$ representing a linear correlation between these two.

Although the number of participating nucleons is not directly measurable experimentally, the corresponding spectator asymmetry arising from the fluctuation in the number of spectators can be accessed through the difference in energy deposited by spectator nucleons in the two Zero-Degree Calorimeters ~\cite{acharya} ,and this is defined as,
\begin{equation}
    \alpha_{spec} = \frac{(N-A) - (N-B)}{(N-A) + (N-B)} = \frac{B-A}{2N - (A+B)}
\end{equation}

where N is the total number of nucleons in each nucleus and (N-A) and (N-B) are the number of spectators.
It can be shown that this spectator asymmetry is related to rapidity shift by,
\begin{equation}
    y_{0} =  \frac{1}{2} ln\frac{(A+B)(1+\alpha_{spec}) - 2N\alpha_{spec}}{(A+B)(1-\alpha_{spec}) + 2N\alpha_{spec}}
\end{equation}

Consequently, the rapidity shift can be experimentally characterized using the spectator asymmetry, which serves as an observable proxy for the participant asymmetry.
In the present work, we have used the AMPT string-melting framework to estimate the $y_{0}$, $\alpha_{part}$ and $\alpha_{spec}$ for Au+Au and Cu+Cu collisions at 200 $A$GeV and 62.4 $A$GeV at RHIC and Deep Learning to predict their correlations using these parameters from AMPT~\cite{ampt}.

The AMPT model is a widely used Monte Carlo transport event generator used to successfully simulate high-energy relativistic heavy-ion collisions both at RHIC and LHC energies. It models complex nuclear interactions by combining initial collision geometry, partonic scattering, hadronization, and final hadronic rescattering. It uses the Heavy Ion Jet Interaction Generator (HIJING) ~\cite{hij1,hij2} to generate initial conditions such as the spatial and momentum distributions of minijet partons and excited soft string and Zhang's Parton Cascade (ZPC) ~\cite{zpc} for partonic scatterings. It converts partons into hadrons via either the Default mode (Lund string fragmentation) or the String Melting mode (quark coalescence). The latter is used to mimic the conditions similar to the de-confined QGP. It utilizes a relativistic transport model (ART) ~\cite{art} to handle final-state interactions and re-scattering among generated hadrons.

\section{Deep Learning Model}


\subsection{Network Architecture}

\begin{figure}[htbp]
    \centering
    \includegraphics[width=\columnwidth]{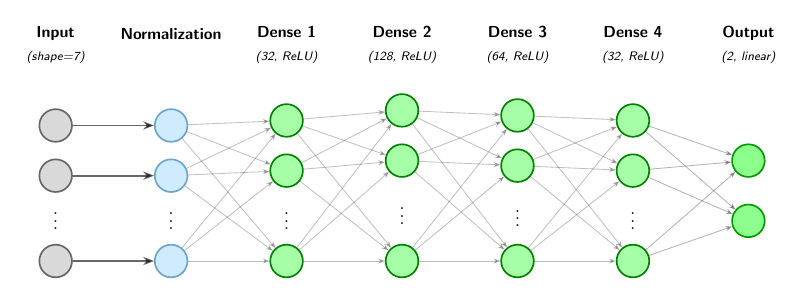}
    \caption{\justifying Schematic of the deep neural network architecture used in this work.}
    \label{fig:schematic_MODEL}
\end{figure}

The regression framework is constructed as a fully connected feed-forward neural network in TensorFlow/Keras, consisting of four hidden layers and a two-node linear output layer (FIG.~\ref{fig:schematic_MODEL}). The model accepts six input features: particle rapidity ($y$), particle multiplicity per unit rapidity ($dN/dy$), spectator asymmetry ($\alpha_{\text{spec}}$), charged-particle multiplicity, beam energy ($\sqrt{s_{NN}}$), and the collision systems, where categorical system identifiers are converted using one-hot encoding. By utilizing experimentally accessible observable parameters as inputs, the model reconstructs initial-state properties which cannot be measured experimentally, specifically the rapidity shift $y_0$ and participant asymmetry $\alpha_{\text{part}}$.

The dataset is organized on a per-event basis, with 80\% of the generated samples allocated for model training and the remaining 20\% reserved for validation and testing. The first layer after the raw 7-dimensional input vector is a \texttt{tf.keras.layers.Normalization} layer, whose feature-wise mean and variance are computed via the \texttt{adapt()} method ~\cite{tensorflow2015} on the training set prior to optimization. Embedding normalization directly within the computational graph ensures that common scale-and-shift normalization is consistently applied during both training and inference without requiring separate tracking of feature scaling statistics. 

The hidden stack adopts a non-monotonic width configuration ($32 \to 128 \to 64 \to 32$), widening in the second layer before narrowing. This allows the network to expand the effective feature space in an intermediate representation before compressing toward the two-dimensional output target. Rectified Linear Units (ReLU) ~\cite{nair2010rectified} are selected as hidden activations to introduce non-linearity, enabling the network to model complex interactions between final-state observables and initial-state geometry while avoiding vanishing gradient issues and minimizing computational overhead. The output layer comprises two linear units with no activation function, providing unbounded, continuous estimates for each regression target. 

Network weights are optimized over 20 training epochs using the Adam optimizer ~\cite{kingma2015adam} with a learning rate of $1\times10^{-4}$. To maintain structural simplicity and low computational latency, no dropout layers or explicit $L_2$ weight regularization techniques are added, as the model demonstrated smooth convergence without empirical evidence of overfitting. The DNN model was developed in Python using the Keras v3.13.2 deep learning API ~\cite{keras}, with TensorFlow v2.20.0 as the underlying framework \cite{tensorflow}, in conjunction with the Scikit-learn machine learning library \cite{scikit}.

\subsection{Model Validation}

The target variable $\alpha_{part}$ exhibits heteroscedastic variance, with the spread of asymmetry values increasing systematically from central to peripheral collisions. To mitigate the impact of this non-uniform variance during training, we adopted the mean absolute error (MAE) as the training loss, given its linear penalization of errors, while additionally monitoring the mean squared error (MSE) and root mean squared error (RMSE) as evaluation metrics. Model performance was tracked via the training and validation MAE curves obtained from the Keras \texttt{History} object. Since the test split was constructed at the event level and sampled independently from each of the twelve source files, the validation loss reflects the model's generalization across collision systems, beam energy, and centrality simultaneously, rather than within a single homogeneous subsample.

\begin{figure}[htbp]
    \centering
    \includegraphics[width=\columnwidth]{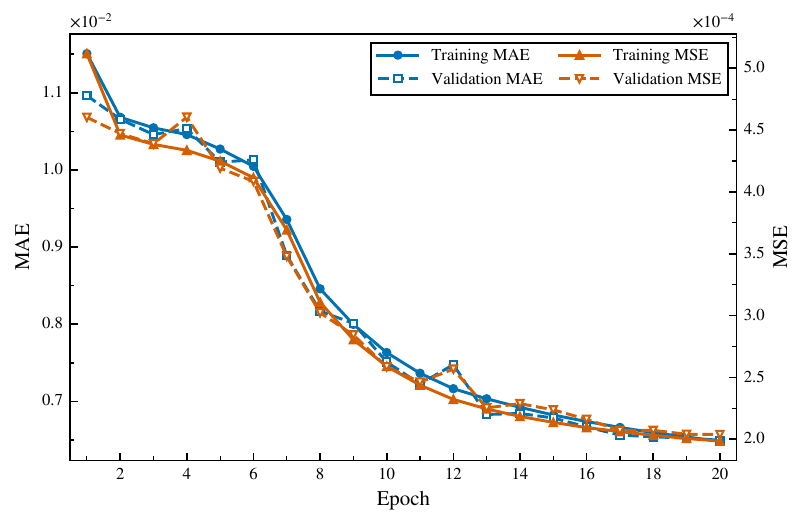}
    \caption{\justifying Epoch-wise training and validation loss measured in MAE and MSE using Adam optimization.}
    \label{fig:loss}
\end{figure}

Figure~\ref{fig:loss} presents the evolution of the training and validation MAE and MSE losses across 20 epochs. At the first epoch, the training losses are approximately $1.15\times10^{-2}$ (MAE) and $5.1\times10^{-4}$ (MSE), while the corresponding validation values are about $1.10\times10^{-2}$ and $4.6\times10^{-4}$. Both metrics decrease steadily, and from epoch 8 onward the training and validation curves nearly coincide, varying only within a narrow range before settling at an MAE of roughly $0.65\times10^{-2}$ and an MSE of roughly $2.0\times10^{-4}$. The small difference between the two sets of curves points to good generalization, with no evidence of overfitting within the 20-epoch training budget. The quality of the predictions for both target quantities was assessed using the coefficient of determination ($R^2$), expressed as follows:


\begin{equation}
    R^2 = 1 - \frac{\sum_{i=1}^{n} (y_i - \hat{y}_i)^2}{\sum_{i=1}^{n} (y_i - \bar{y})^2}
    \label{r2_eqn}
\end{equation}

In Eq. \ref{r2_eqn}, $y_i$ represents the actual target value, $\hat{y}_i$ the predicted value, and $\bar{y}$ the mean of the actual values across $n$ samples. 

Table~\ref{tab:r2_score} reports the average value of the coefficient of determination for the two regression targets, showing good agreement between the predicted and the true values for both $y_0$ and $\alpha_{part}$. The comparable $R^2$ values across both training and test data demonstrate that the model generalizes well without overfitting. These near-unity scores indicate that the model explains nearly all of the variance in both targets, consistent with the low MAE loss shown in Figure \ref{fig:loss}.

\begin{table}[h]
\centering
\caption{\justifying Average coefficient of determination ($R^2$) score for the two regression targets on train and test datasets}
\begin{tabular}{l l l}
\hline 
\multirow{2}{*}{Variable} & \multicolumn{2}{c}{$R^2$ score} \\
 & Train dataset & Test dataset \\
\hline
$y_0$           & 0.98658 & 0.98600 \\
$\alpha_{part}$ & 0.98892 & 0.98851 \\
\hline 
\end{tabular}
\label{tab:r2_score}
\end{table}

\section{Model Performance}

\begin{figure}[htbp]
    \centering
    \includegraphics[width=\columnwidth]{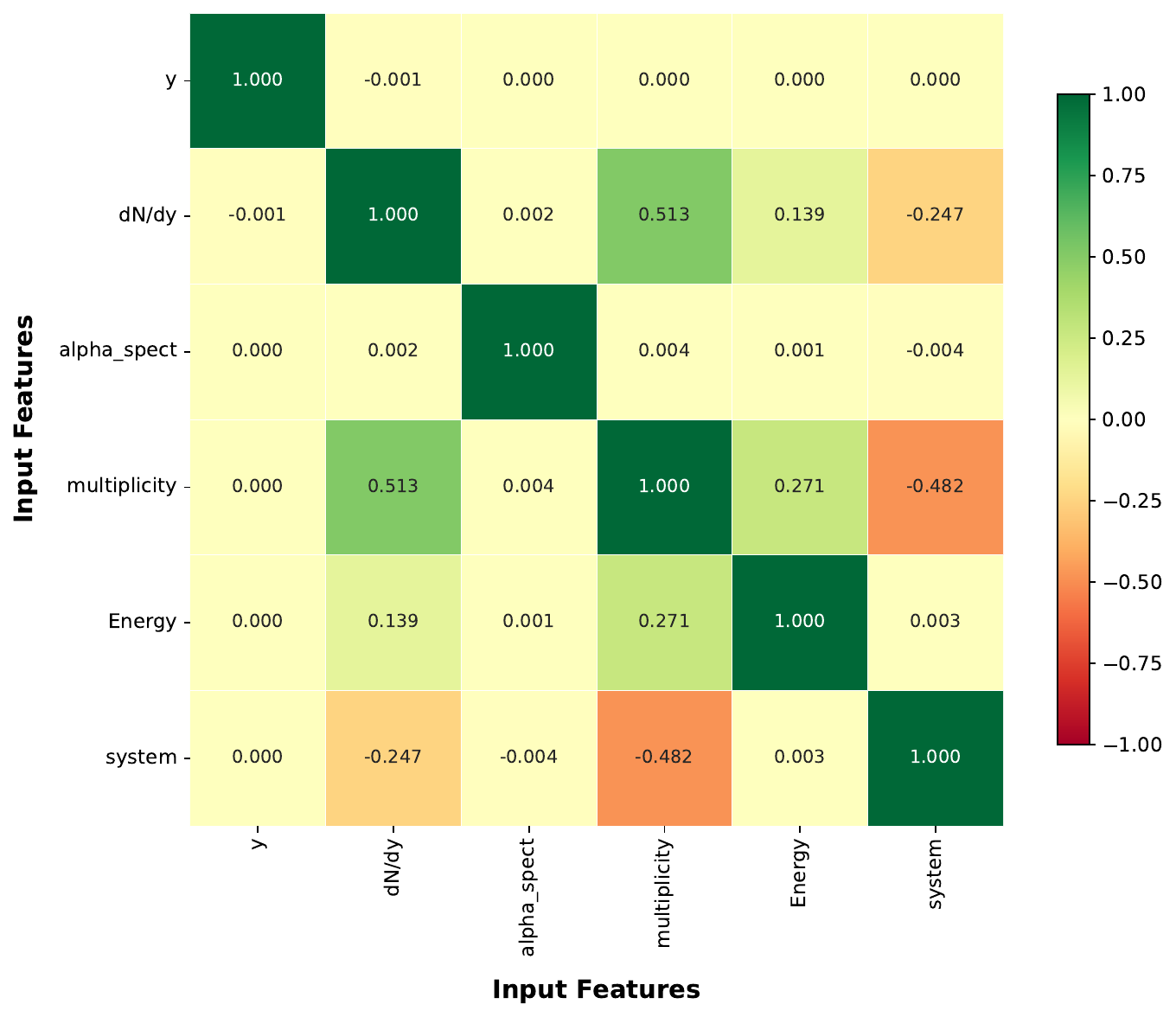}
    \caption{\justifying Pearson correlation heatmap of the six input features utilized in model training.}
    \label{fig:Input_Features_Correlation}
\end{figure}

Model fidelity was evaluated by combining Pearson correlation with  SHAP, allowing raw data correlations to be disentangled from learned feature importances. The Pearson matrix captures linear relationships between input attributes, while SHAP analysis isolates individual feature impacts despite feature interactions and correlations. The Pearson correlation matrix shown in Figure ~\ref{fig:Input_Features_Correlation} illustrates the pairwise linear relationships among the dataset's input features. 
Most pairs are essentially uncorrelated ($ < 0.01$), and the only moderate correlations occur among the event-activity and system variables: dN/dy with multiplicity (0.513), multiplicity with energy (0.271), and the system variable with multiplicity ($-$0.482) and dN/dy ($-$0.247), consistent with higher particle production at higher energy and in the larger Au+Au system. Since no pair exceeds  0.52, the input set shows no strong multicollinearity, indicating that the learned dependencies are not mainly driven by redundant features.


\begin{figure}[htbp]
    \centering
    \includegraphics[width=\columnwidth]{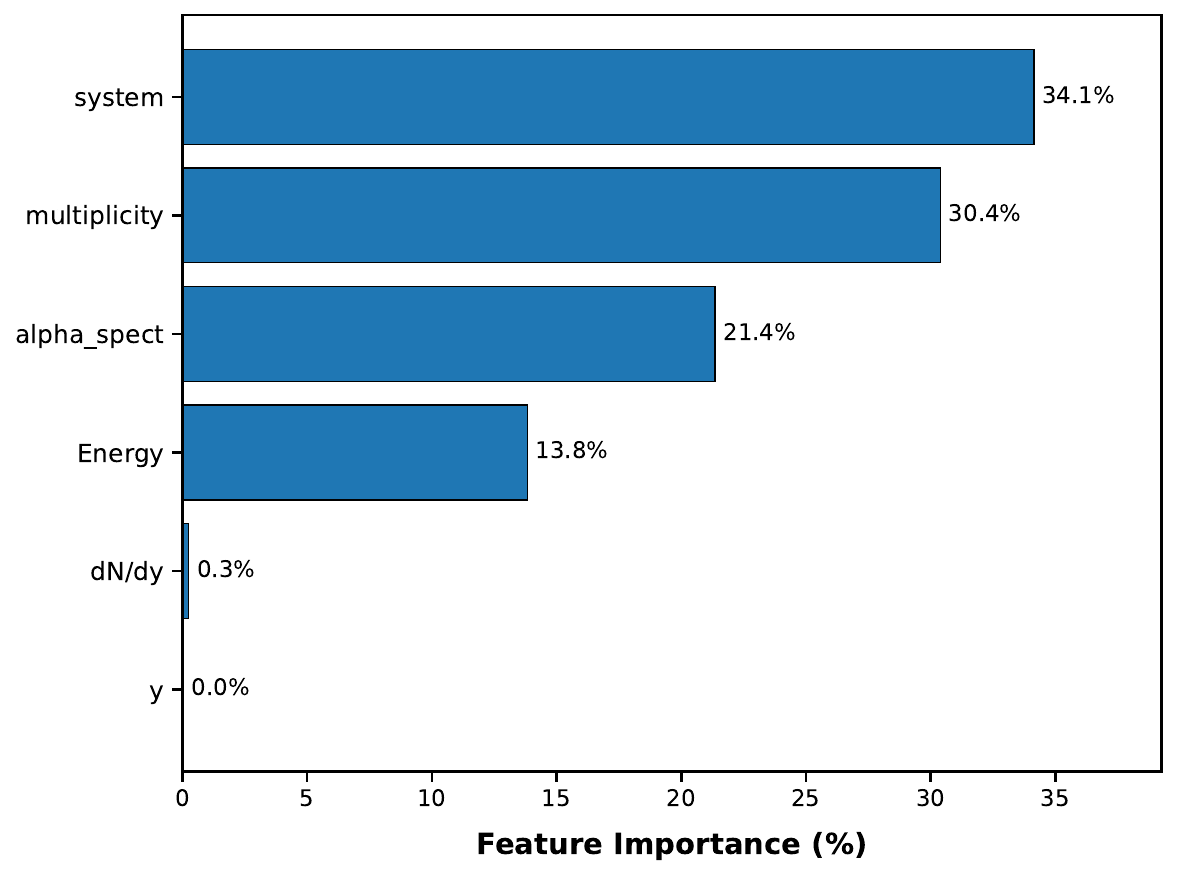}
    \caption{\justifying Feature importance determined by SHAP analysis, showing the contribution of each input variable to model output.}
    \label{fig:shap}
\end{figure}


Figure~\ref{fig:shap} shows that the model’s predictions are dominated by the collision system (34.1\%) and multiplicity (30.4\%), followed by $\alpha_{spect}$ (21.4\%) and collision energy (13.8\%), while dN/dy (0.3\%) and y (0.0\%) contribute negligibly. The substantial weight on $\alpha_{spect}$, despite its negligible linear correlation with the other inputs (FIG. ~\ref{fig:Input_Features_Correlation}), indicates a genuine nonlinear dependence learned by the network rather than redundancy among features. The low importance of dN/dy, which is moderately correlated with multiplicity, suggests that the model draws the shared information from multiplicity and treats dN/dy as largely redundant.

\begin{figure}[htbp]
    \centering
    \includegraphics[width=\columnwidth]{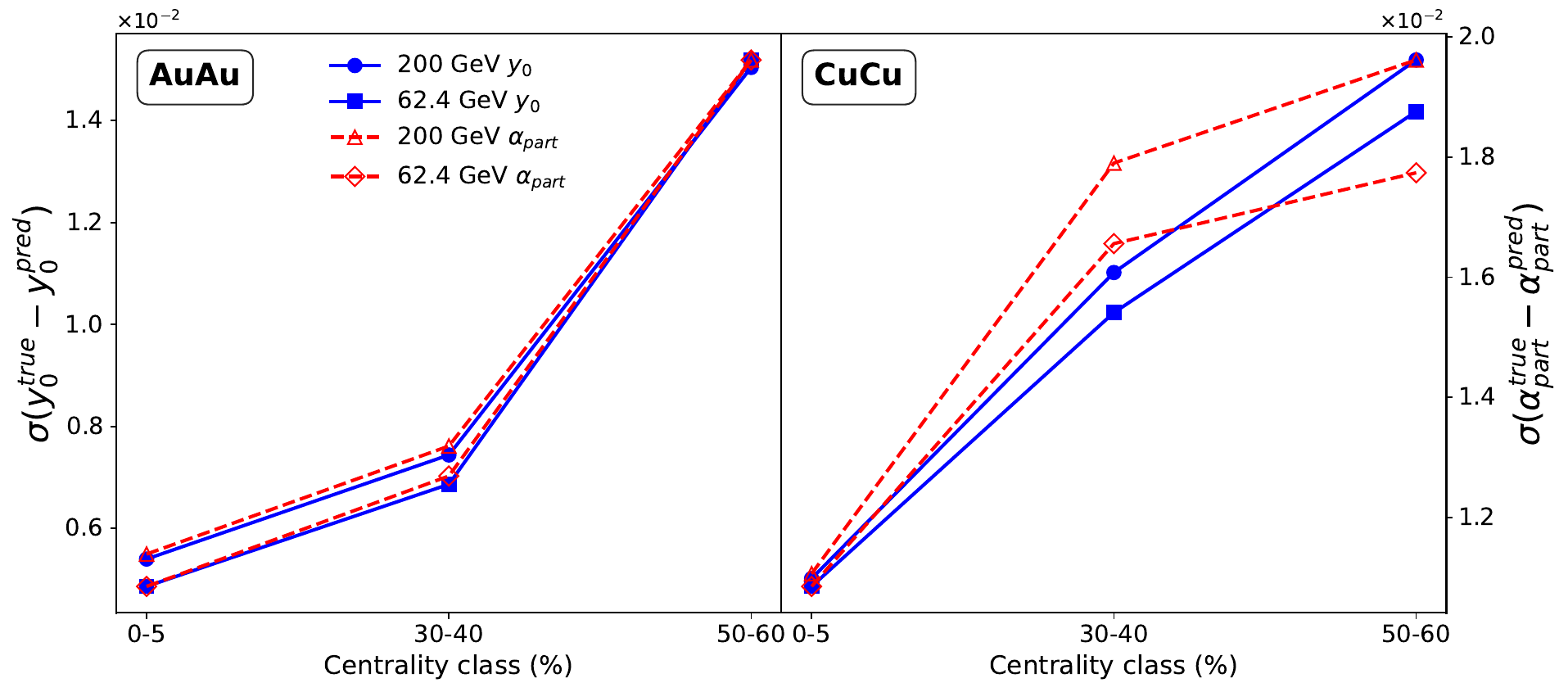}
    \caption{\justifying Resolution of the model predictions as a function of centrality class for Au+Au (left) and Cu+Cu (right) collisions at $\sqrt{s_{NN}}$ = 62.4 and 200 GeV. For each system, energy, and centrality class, the residual distribution (true − predicted) of $y_{0}$  and $\alpha_{part}$  was fitted with a Gaussian, and the fitted width ($\sigma$) is plotted.}
    \label{fig:resolution}
\end{figure}


To quantify the precision of the predictions, we fitted the residual distributions (true $-$ predicted) of $y_0$ and $\alpha_{part}$ with a Gaussian function for each collision system, beam energy, and centrality class. The fitted mean values are close to zero, indicating that the model has no significant systematic bias, and the fitted width ($\sigma$) is taken as a measure of the prediction resolution. Figure \ref{fig:resolution} shows $\sigma$ as a function of centrality. The resolution is of order $10^{-2}$ and degrades from central to peripheral collisions, which is consistent with larger event-by-event fluctuations for smaller participant numbers. The Au+Au results are nearly independent of beam energy, while Cu+Cu shows a modest energy dependence in the mid-central and peripheral bins.

\section{Results and Discussions}

In experimental analyses, assigning events to a centrality class through the impact parameter is not straightforward, as centrality bins are determined a posteriori using the charged particle multiplicity distribution. In the present exploratory study, we adopt a fixed impact parameter approach to represent different centrality classes in order to reduce the computational cost. Monte Carlo transport models such as AMPT provide a controlled setting in which the true impact parameter is known, making them well suited for training supervised machine-learning regressors that can subsequently be applied. We adopt the impact-parameter range corresponding to each centrality class from Ref.~\cite{impact1,impact2}.



\begin{figure}[htbp]
    \centering
    \includegraphics[width=\columnwidth]{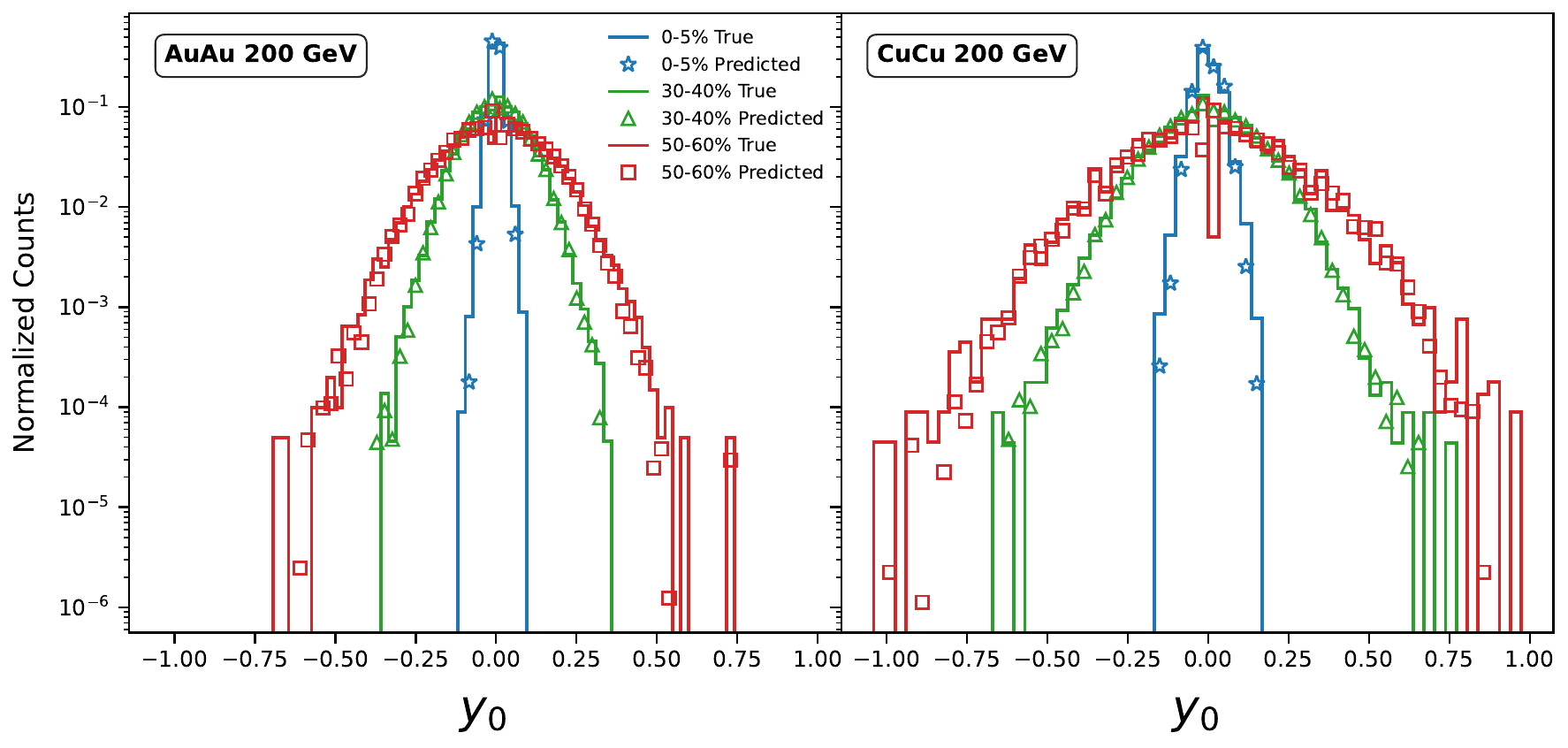}
    \caption{\justifying True and predicted distributions of $y_0$ across three centrality classes (0–5\%, 30–40\%, and 50–60\%) for Au+Au (left panel) and Cu+Cu (right panel) collisions at $\sqrt{s_{NN}}$ = 200 $A$\text{GeV}.}
    \label{fig:y0_200}
\end{figure}

Figure \ref{fig:y0_200} shows the event-by-event distribution of $y_0$ for Cu+Cu and Au+Au collisions at 200 GeV in three centrality bins using a Monte Carlo  Glauber initial state ~\cite{mcg1,mcg2,mcg3,mcg4}.
 
It compares the ground-truth and predicted distributions of the observable $y_0$ across three centrality classes (0–5\%, 30–40\%, and 50–60\%) for Au+Au and Cu+Cu collisions at $\sqrt{s_{NN}} = 200\text{ GeV}$. Across all centrality bins and both colliding systems, the predicted values exhibit good agreement with the true normalized distributions across several orders of magnitude, accurately capturing both the peak structures and the low-probability tails. Central collisions (0–5\%) yield narrow, sharply peaked distributions around $y_0 = 0$, whereas mid-central and peripheral events exhibit progressively broader distributions, extending up to $y_0 \approx \pm 0.7$ for Au+Au and $y_0 \approx \pm 1.0$ for Cu+Cu at 50–60\% centrality. The width of the $y_0$ distribution increases from central to peripheral collisions as the relative fluctuation increases for larger impact parameters, as the number of participants is relatively smaller for those collisions. Furthermore, the lighter Cu+Cu system consistently displays wider distribution profiles than Au+Au for equivalent centrality intervals, reflecting larger geometric variations in smaller collision volumes. Overall, the close overlap between the true and reconstructed distributions demonstrates that the model faithfully predicts $y_0$ without introducing systematic bias across varying system sizes and centralities.

\begin{figure}[htbp]
    \centering
    \includegraphics[width=0.5\textwidth]{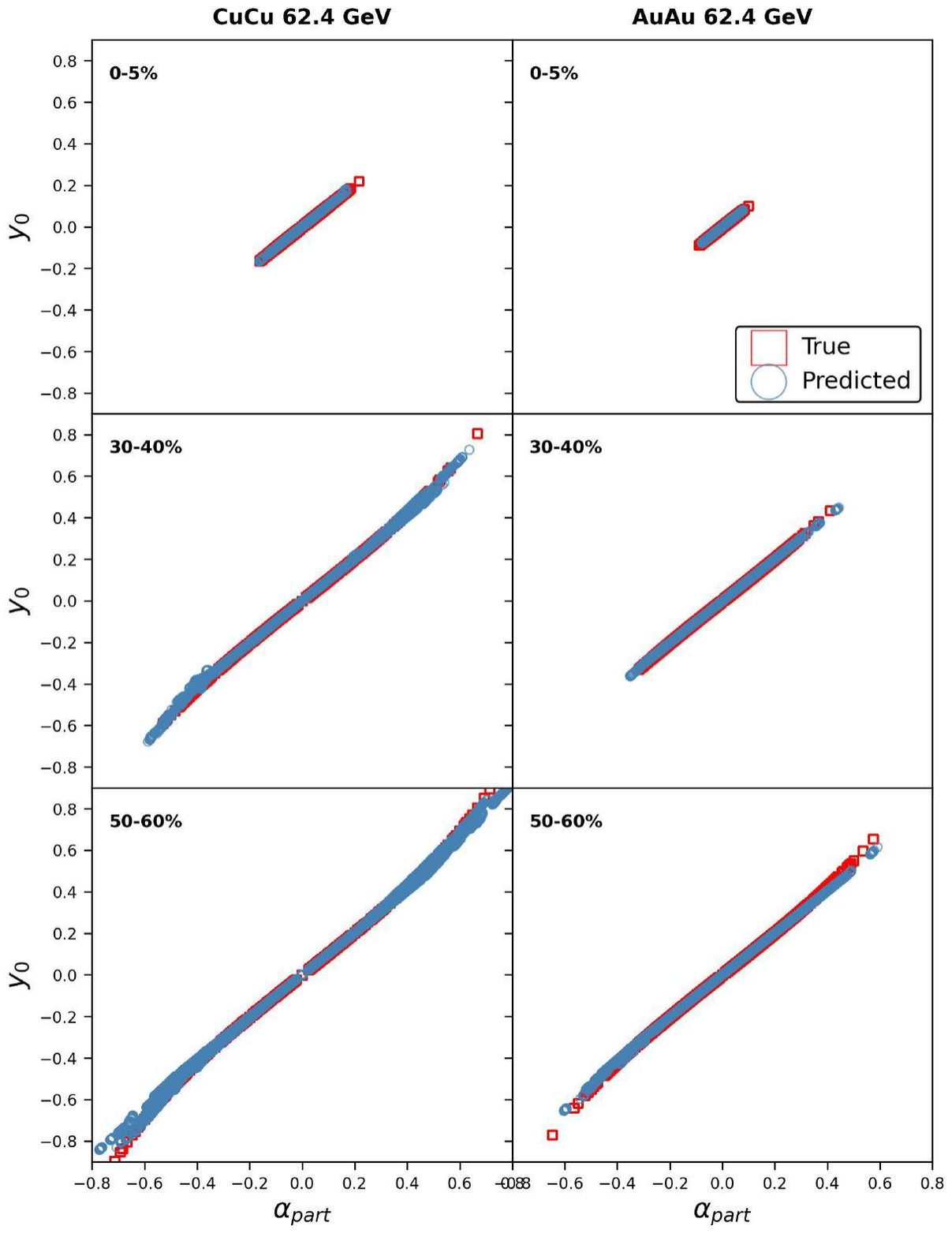}
    \caption{\justifying Correlation between target observables $y_0$ and $\alpha_{part}$, comparing true values (red) and model predictions (steelblue) across 0–5\%, 30–40\%, and 50–60\% centrality classes for Cu+Cu (left) and Au+Au (right) collisions at $\sqrt{s_{NN}}$  = 62.4 $A$\text{GeV}.}
    \label{fig:y0_vs_alphapart_62p4}
\end{figure}

\begin{figure}[htbp]
    \centering
    \includegraphics[width=0.5\textwidth]{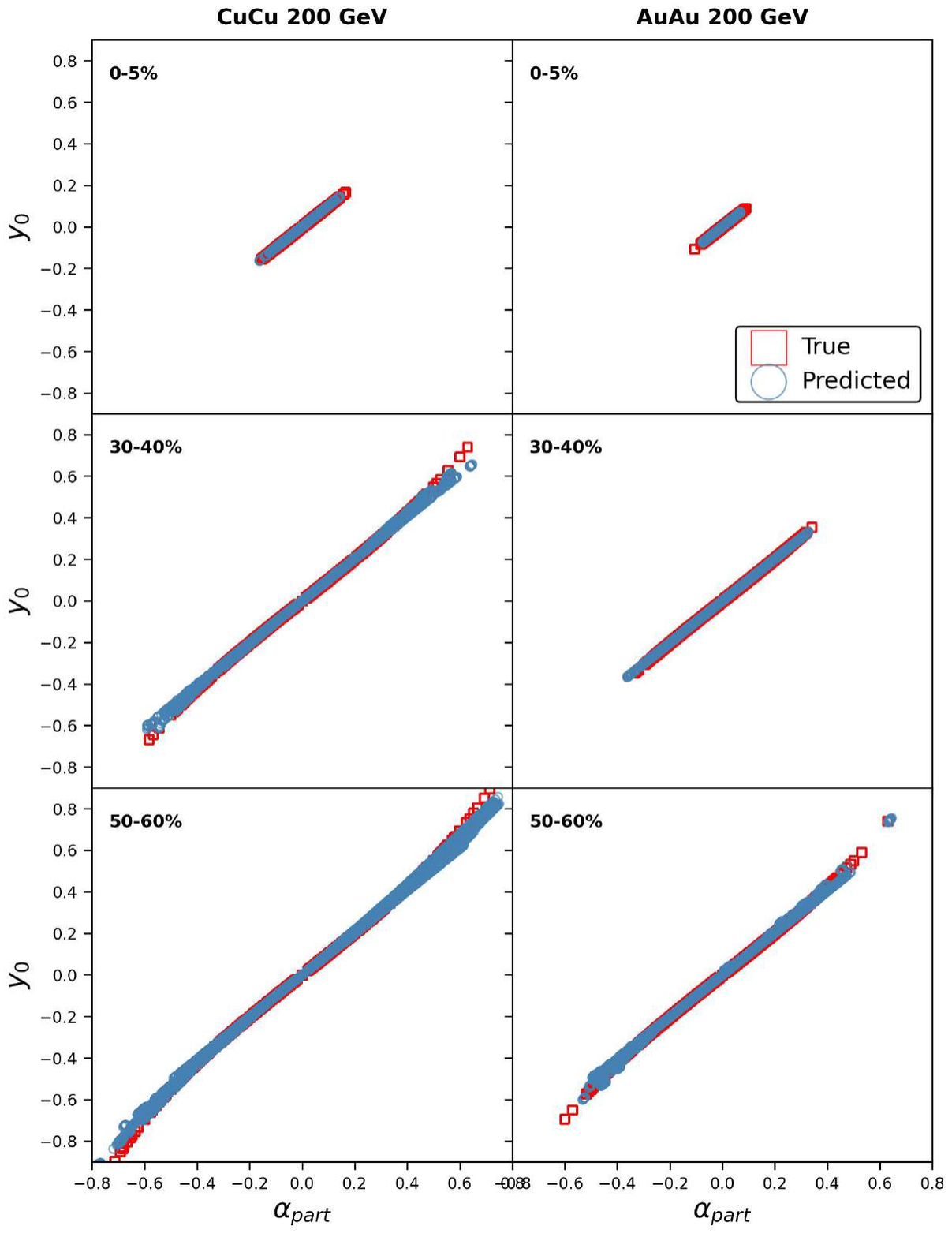}
    \caption{\justifying Correlation between target observables $y_0$ and $\alpha_{part}$ comparing true values (red) and model predictions (steelblue) across 0–5\%, 30–40\%, and 50–60\% centrality classes for Cu+Cu (left) and Au+Au (right) collisions at $\sqrt{s_{NN}} = 200\text{ AGeV}$.}
    \label{fig:y0_vs_alphapart}
\end{figure}

\subsection{Predictive performance on $y_0$ vs $\alpha_{part}$}

\begin{figure}[t]
    \includegraphics[width=0.5\textwidth]{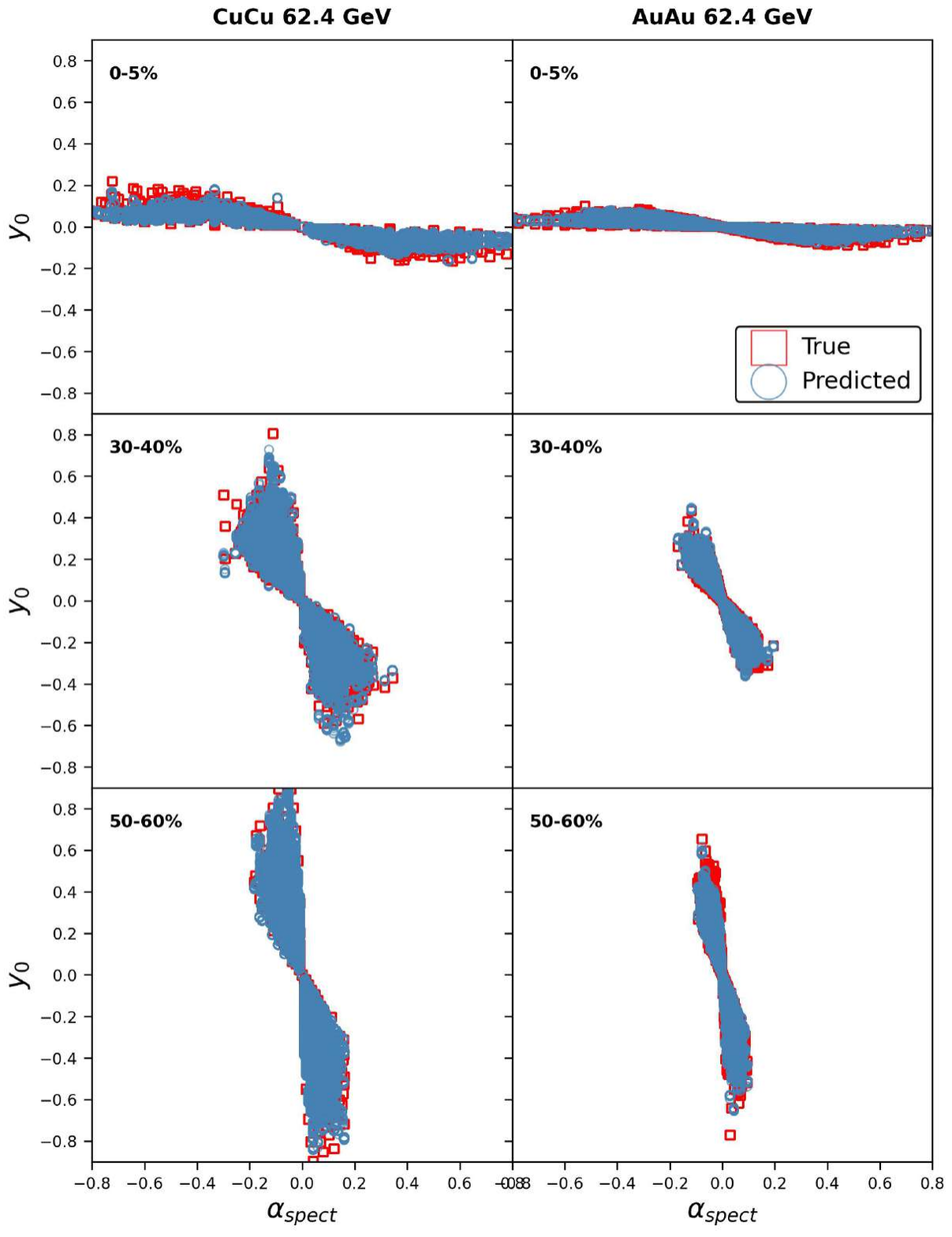}
    \caption{\justifying Correlation between $y_0$ and $\alpha_{spect}$ comparing true values (red) and model predictions (steelblue) across 0–5\%, 30–40\%, and 50–60\% centrality classes for Cu+Cu (left) and Au+Au (right) collisions at $\sqrt{s_{NN}} = 62.4\text{ AGeV}$.}
    \label{fig:y0_vs_alphaspect}
\end{figure}

\begin{figure}[t]
    \centering
    \includegraphics[width=0.5\textwidth]{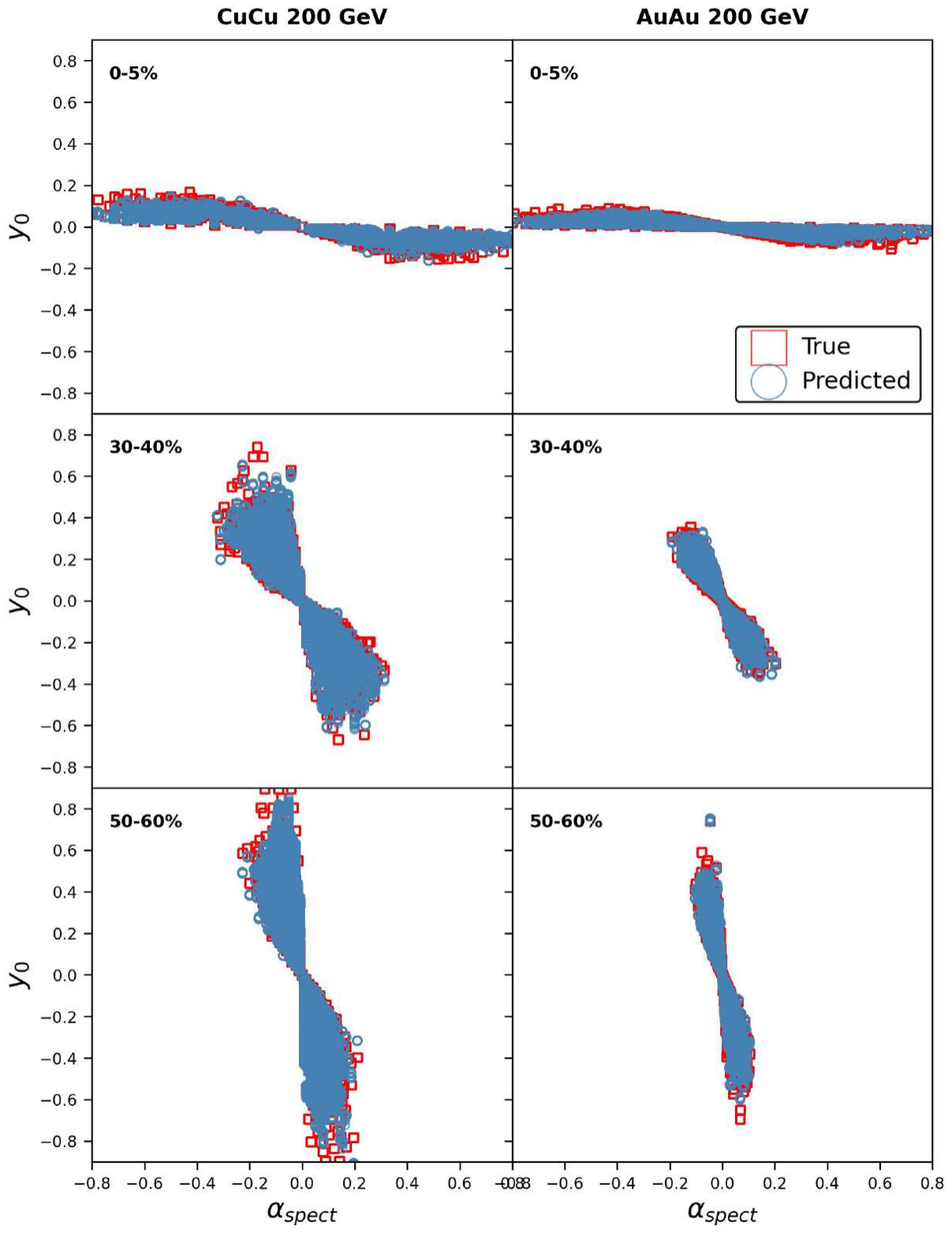}
    \caption{\justifying Correlation between $y_0$ and $\alpha_{spect}$ comparing true values (orange) and model predictions (blue) across 0–5\%, 30–40\%, and 50–60\% centrality classes for Cu+Cu (left) and Au+Au (right) collisions at $\sqrt{s_{NN}} = $200 AGeV.}
    \label{fig:y0_vs_alphaspect_200}
\end{figure}

Figure \ref{fig:y0_vs_alphapart_62p4} and Figure~\ref{fig:y0_vs_alphapart} present the phase-space correlation between the target observables $y_0$ and $\alpha_{part}$, comparing true values against model predictions for Cu+Cu and Au+Au collisions at $\sqrt{s_{NN}} = 62.4\text{ GeV}$ and 200 GeV respectively across three centrality classes: 0–5\%, 30–40\%, and 50–60\%. Across all centralities and collision systems, $y_0$ and $\alpha_{part}$  exhibit a strong, almost linear underlying correlation. The model predictions show excellent agreement with the ground truth, completely capturing the continuous phase-space distribution without introducing artificial dispersion or systematic bias. The dynamic range of both observables expands systematically from central to peripheral collisions; central events (0–5\%) remain tightly localized near the origin ($y_0 , \alpha_{part} \in [-0.2, 0.2]$), whereas peripheral events (50–60\%) span a significantly broader range ($\sim [-0.8, 0.6]$). Additionally, the smaller Cu+Cu system exhibits a wider phase-space coverage compared to Au+Au within the same centrality intervals, consistent with increased geometry-driven event-by-event fluctuations in smaller interaction volumes. The predicted points reproduce this near-linear relationship with high fidelity across all nine panels, with only minor scatter appearing at the extreme edges of the 50–60\% centrality bin, reinforcing that prediction accuracy for both targets jointly degrades mildly toward peripheral events while remaining physically consistent throughout.

\subsection{Predictive performance on $y_0$ vs $\alpha_{spect}$}
However, such a strong linear correlation is not observed in the distribution of $\alpha_{spec}$ and $y_0$ shown in Figure~\ref{fig:y0_vs_alphaspect} and \ref{fig:y0_vs_alphaspect_200} due to the presence of the (A+B) term in Eq. (6). There is a distribution of $y_0$ for a given value of $\alpha_{spec}$. The spectator asymmetry is maximum for central collisions due to a very small number of spectators and least for most peripheral collisions, which is reflected by the slopes of the distributions. This centrality-dependent change in the $y_0$ –$\alpha_{spec}$ relationship is reproduced almost exactly by the model's predictions in all three collision systems, indicating that the network captures the underlying physical correlation structure rather than fitting each target independently.

The rapidity shift $y_0$ has a direct impact on the final state charged particles' rapidity distributions. This effect has been studied in Pb+Pb collisions at 2.76 $A$TeV from the measurement of spectator energy deposition at
the ZDCs by the ALICE Collaboration ~\cite{acharya} and in Pb+Pb, Au+Au, and Cu+Cu using model frameworks ~\cite{raniwala1, Thakur}. This effect was studied by taking the ratio of the dN/dy of symmetric and asymmetrical events, which can be described by a Taylor series expansion ~\cite{acharya,raniwala1, Thakur}. The coefficients of the expansion depend on the shape of the actual rapidity distribution and rapidity shift. The linear coefficient is related to the rapidity shift by $c_1 = \frac{y_0}{\sigma^2}$ where $\sigma$ is the width of the rapidity distribution, $c_2 \propto y_0^2$, and so on. The magnitudes of the coefficients signify the magnitudes of asymmetry, confirming that initial state longitudinal asymmetry can affect the final state observables.

\begin{figure}[htbp]
    \centering
    \begin{subfigure}[b]{\columnwidth}
        \centering
        \includegraphics[width=\columnwidth]{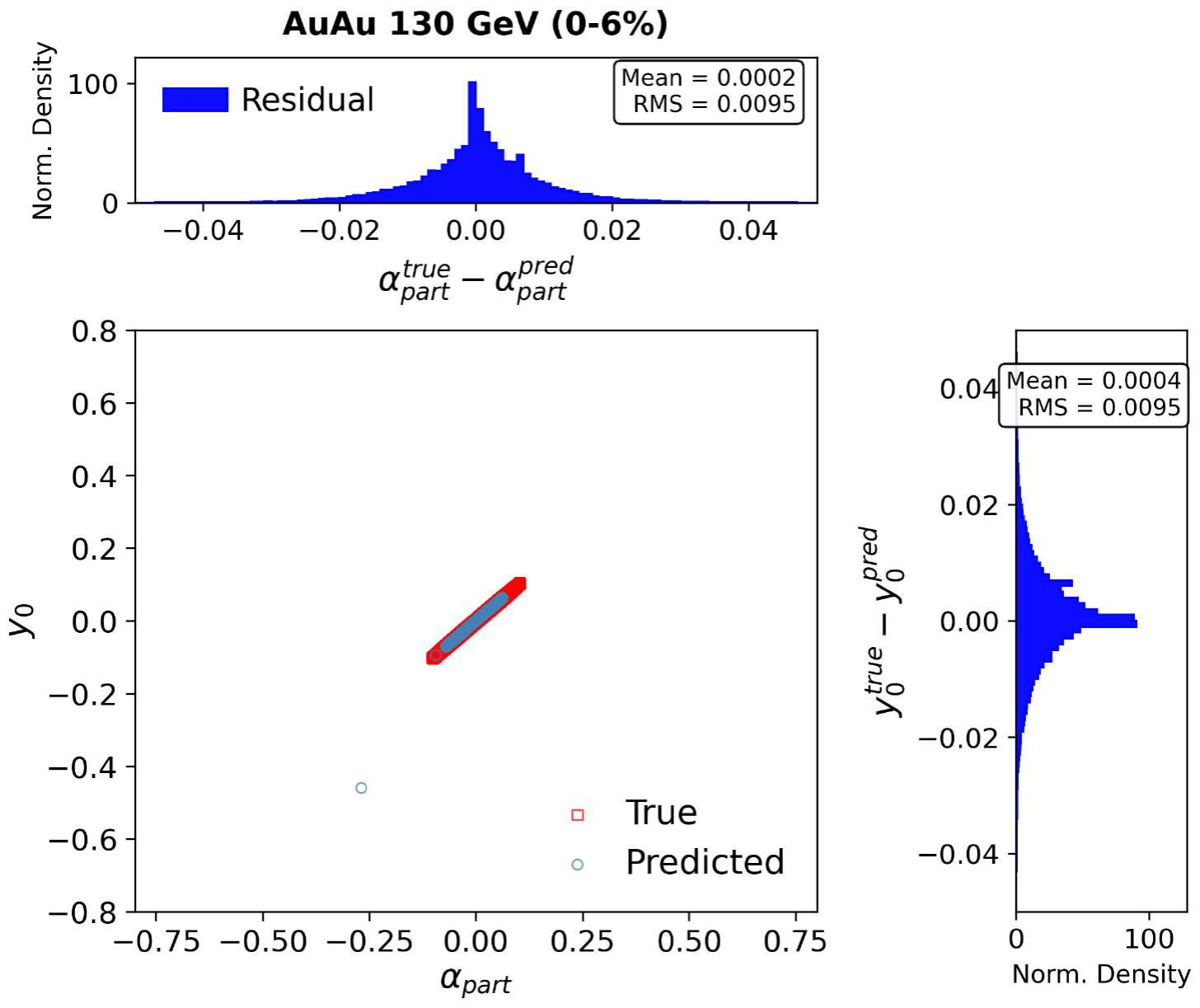}
        \caption{$0\text{--}6\%$}
        \label{fig:y0_vs_alpha_part_06}
    \end{subfigure}

    \begin{subfigure}[b]{\columnwidth}
        \centering
        \includegraphics[ width=\columnwidth]{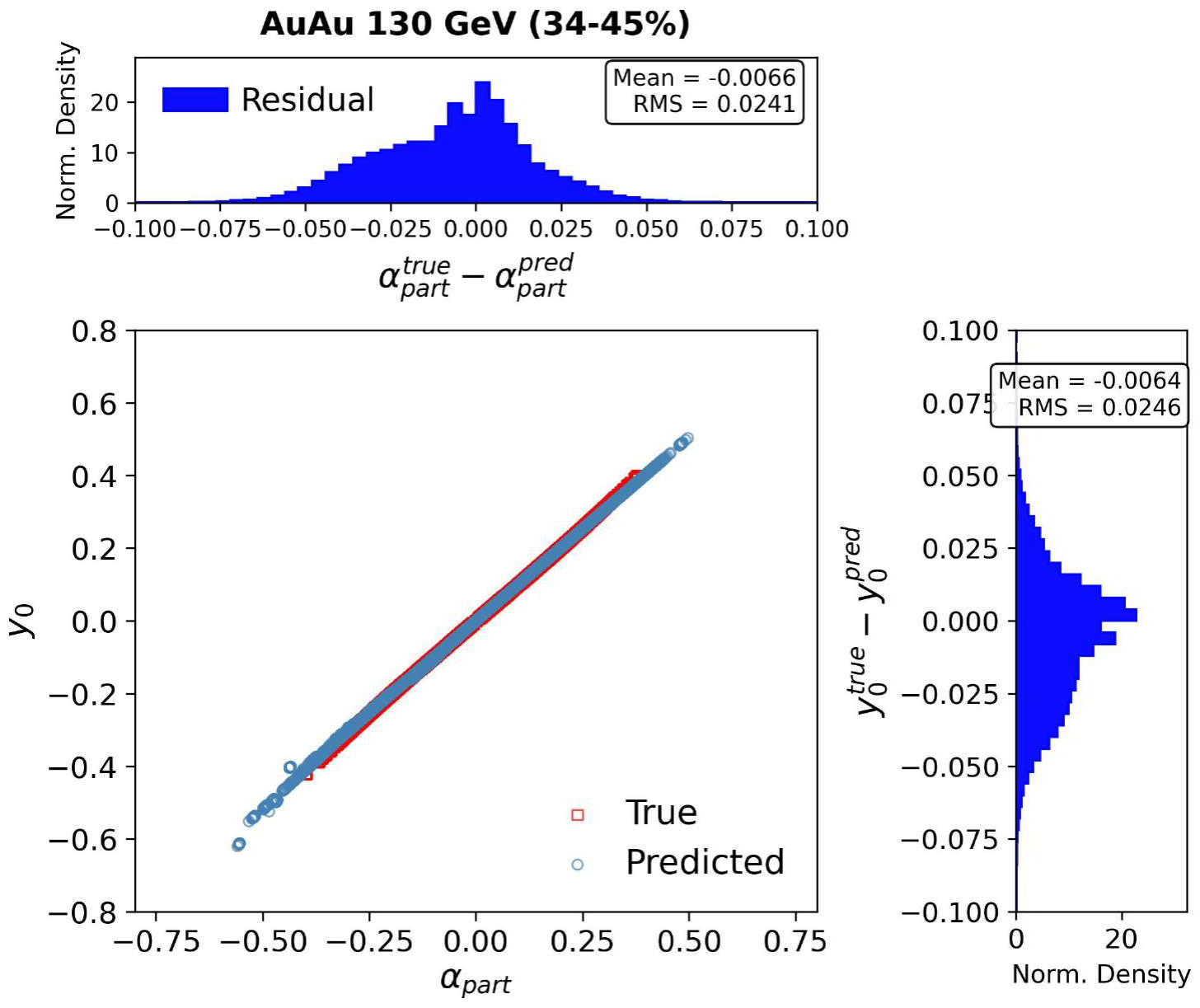}
        \caption{$34\text{--}45\%$}
        \label{fig:y0_vs_alpha_part_3445}
    \end{subfigure}

    \begin{subfigure}[b]{\columnwidth}
        \centering
        \includegraphics[ width=\columnwidth]{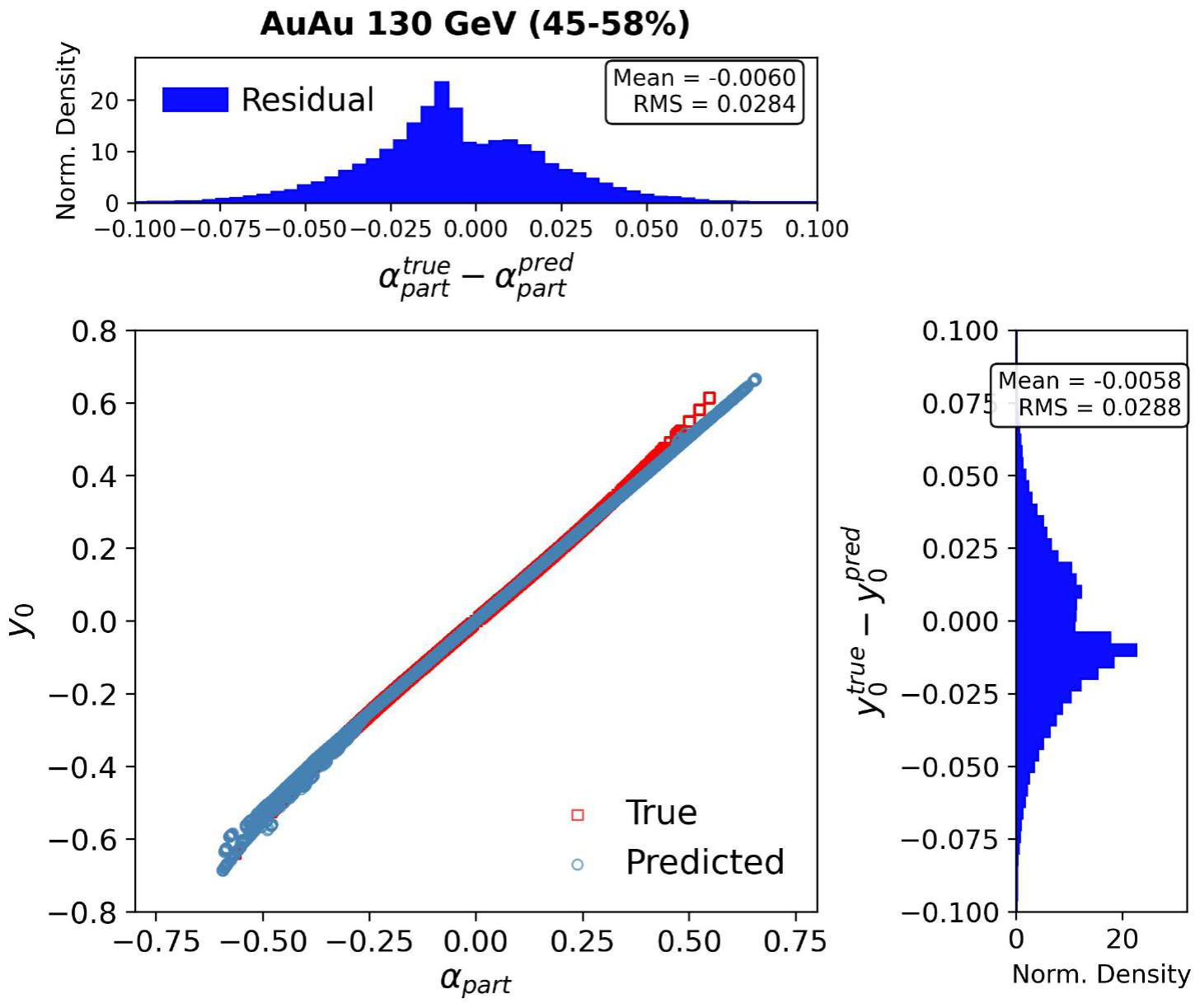}
        \caption{$45\text{--}58\%$}
        \label{fig:y0_vs_alpha_part_4558}
    \end{subfigure}

\caption{\justifying Phase-space correlation between $y_0$ and $\alpha_{part}$ in the unseen centrality regions: (a) $0\text{--}6\%$, (b) $34\text{--}45\%$, and (c) $45\text{--}58\%$ for Au+Au collisions at $\sqrt{s_{NN}} = 130$ AGeV. Top panel: residual distribution of $\alpha_{part}^{true} - \alpha_{part}^{pred}$. Main panel: true (red) vs.\ predicted (steelblue) values of $y_0$ as a function of $\alpha_{part}$. Right panel: residual distribution of $y_0^{true} - y_0^{pred}$.}

    \label{fig:y0_vs_alpha_part_AuAu_130}
\end{figure}

\begin{figure*}[htbp]
    \centering
    \begin{subfigure}[b]{1.5\columnwidth}
        \centering
        \includegraphics[trim=1pt 100pt 1pt 100pt, clip, width=\columnwidth]{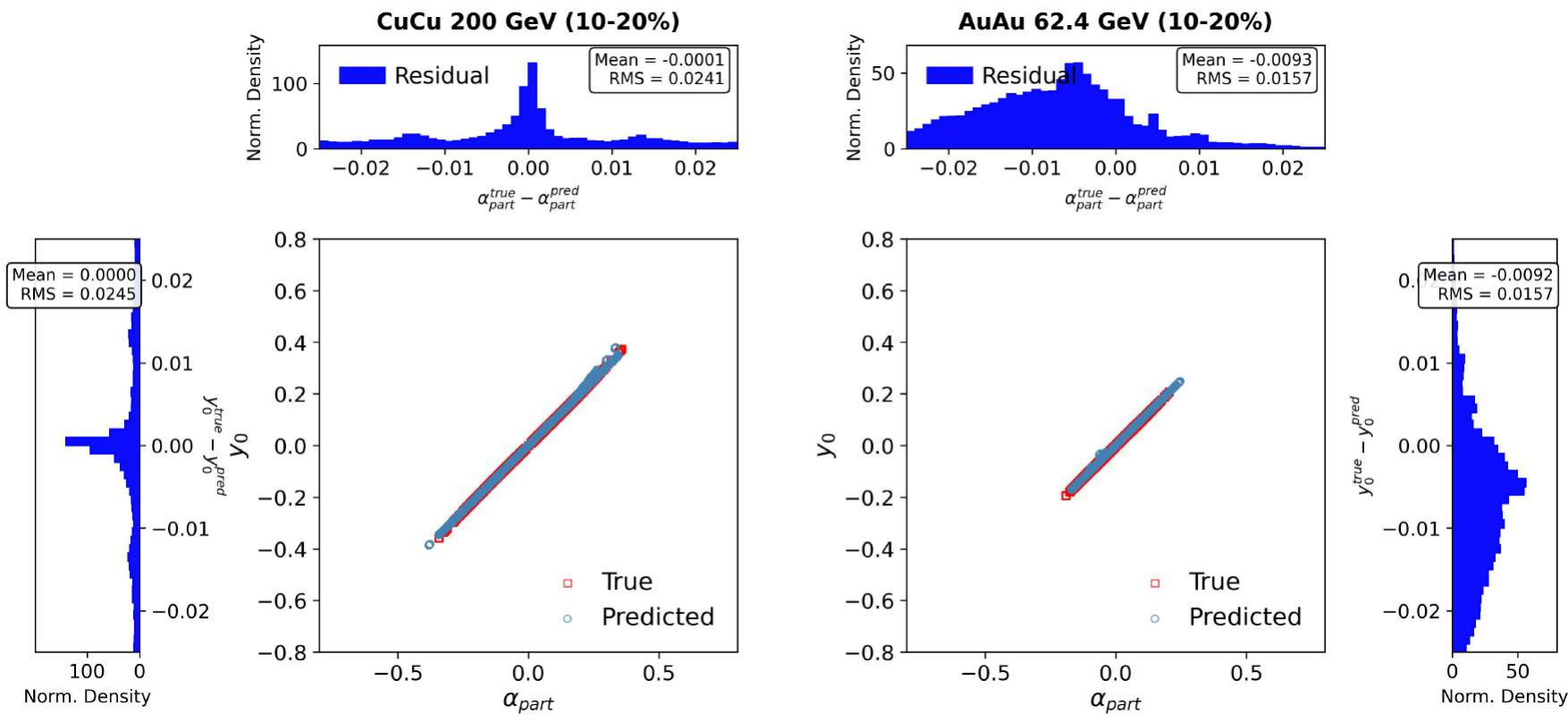}
        \caption{$10\text{--}20\%$}
        \label{fig:y0_1020}
    \end{subfigure}

    \begin{subfigure}[b]{1.5\columnwidth}
        \centering
        \includegraphics[trim=1pt 100pt 1pt 100pt, clip, width=\columnwidth]{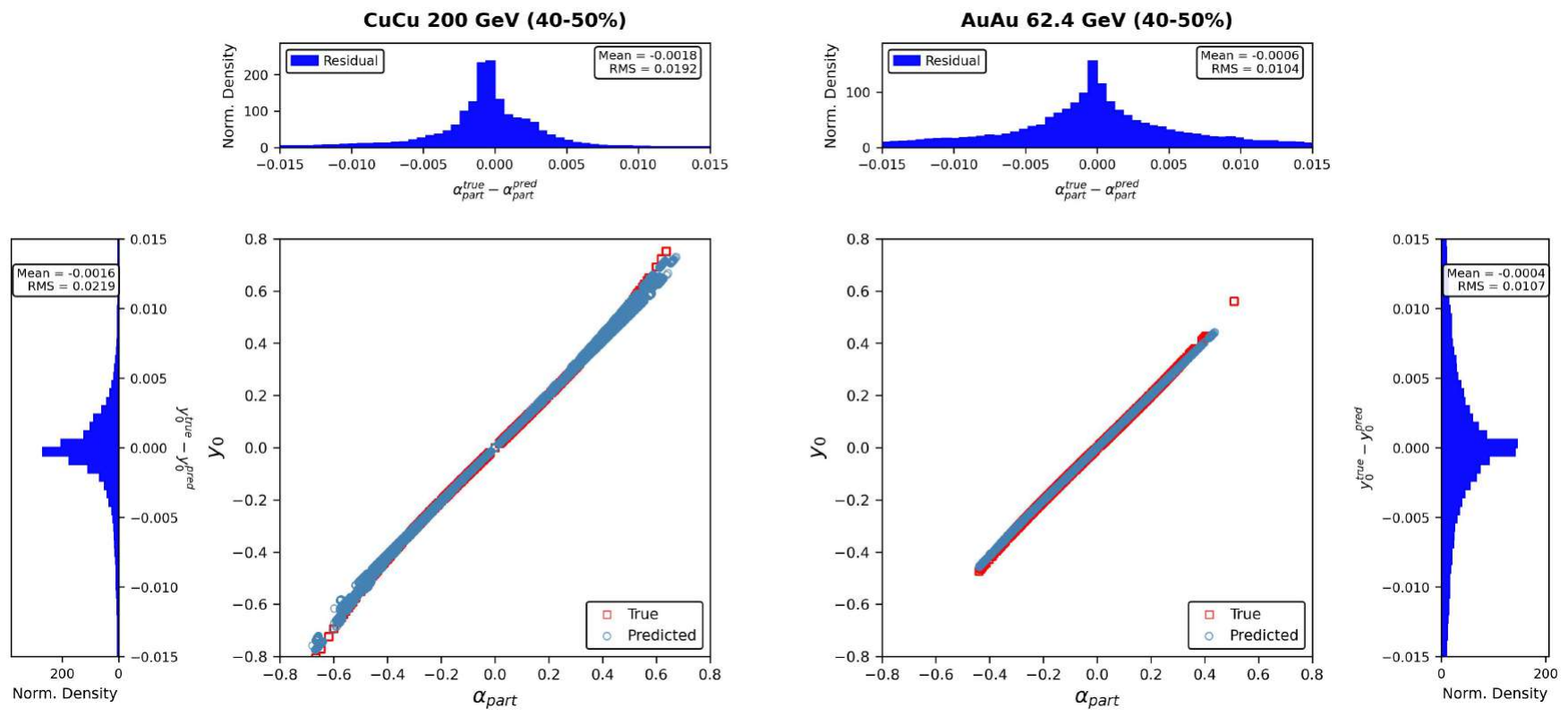}
        \caption{$40\text{--}50\%$}
        \label{fig:y0_4050}
    \end{subfigure}

    \begin{subfigure}[b]{1.5\columnwidth}
        \centering
        \includegraphics[trim=1pt 100pt 1pt 100pt, clip, width=\columnwidth]{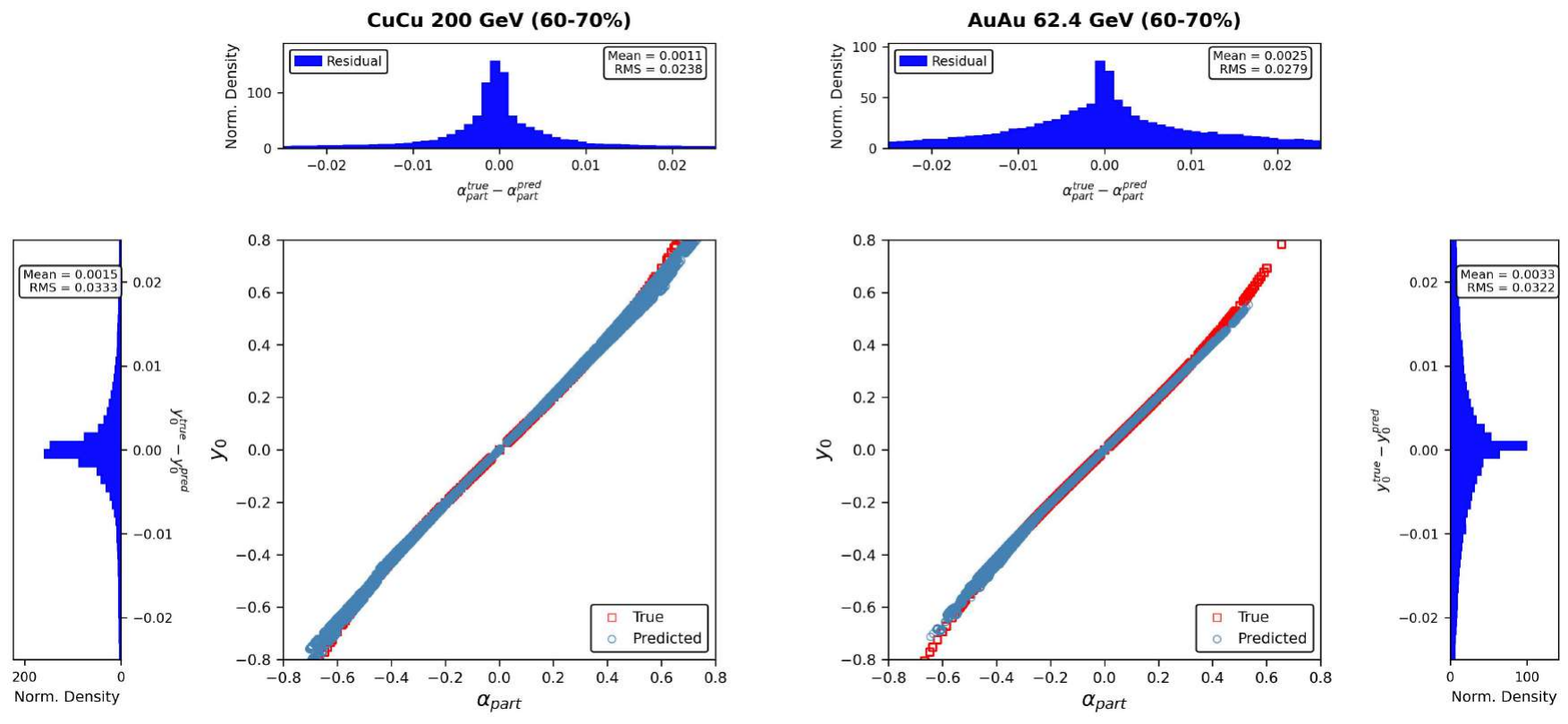}
        \caption{$60\text{--}70\%$}
        \label{fig:y0_6070}
    \end{subfigure}

\caption{\justifying Phase-space correlation between $y_0$ and $\alpha_{part}$ in the unseen centrality regions: (a) $10\text{--}20\%$, (b) $40\text{--}50\%$, and (c) $60\text{--}70\%$ for Cu+Cu at $\sqrt{s_{NN}} = 200\text{ AGeV}$ and Au+Au at $62.4\text{ AGeV}$. Top panel: residual distribution of $\alpha_{part}^{true} - \alpha_{part}^{pred}$. Main panel: true (red) vs.\ predicted (steelblue) values of $y_0$ as a function of $\alpha_{part}$. Right panel: residual distribution of $y_0^{true} - y_0^{pred}$.}
  
    \label{fig:y0_vs_alpha_part_AuAu_CuCu}
\end{figure*}

\subsection{Generalization to unseen centrality intervals}


To evaluate the model's generalization capability on out-of-distribution data, performance was tested on unseen centrality intervals excluded during training: $0\text{--}6\%$, $34\text{--}45\%$, and $45\text{--}58\%$ for Au+Au collisions at $130\text{ GeV}$ (Figs.~\ref{fig:y0_vs_alpha_part_AuAu_130}) and $10\text{--}20\%$, $40\text{--}50\%$, and $60\text{--}70\%$ for Cu+Cu collisions at $200\text{ GeV}$ and Au+Au collisions at $62.4\text{ GeV}$ (Fig.~\ref{fig:y0_vs_alpha_part_AuAu_CuCu}). Despite never encountering these specific centrality intervals during training, the model's predictions align well with the ground-truth values. The model successfully interpolates the non-trivial, centrality-dependent transition in phase space structure evolving from a negatively sloped, broad distribution in central to mid-central events to a highly localized, nearly vertical feature in peripheral collisions. In every case, the predicted $y_0$--$\alpha_{part}$ correlation tracks the true values closely, with residual distributions tightly peaked near zero, confirming that the model generalizes this relationship beyond the collision topologies seen during training.

\begin{table*}[htbp]

\caption{\justifying Quantitative evaluation of $y_0$ and $\alpha_{part}$  predictions on centrality classes not used in training, for Au+Au at 62.4 and 130 GeV and Cu+Cu at 200 GeV. For each system, energy, and centrality bin, the RMS of the true values is compared with the RMS of the residual (true $-$ predicted).}

\label{tab:mean_rms_params}
\begin{tabular}{c c c| c c c c}
 \hline 
\multirow{2}{*}{System} & \multirow{2}{*}{Energy} & \multirow{2}{*}{Centrality} & \multicolumn{4}{c}{RMS }  \\

  & &  & $y_0^{true}$ &   $y_0^{true}$  - $y_0^{pred}$   & $\alpha_{part}^{true}$   & $\alpha_{part}^{true}$ - $\alpha_{part}^{pred}$\\
\hline
\multirow{3}{*}{AuAu}  & \multirow{3}{*}{62.4} & 10-20 &   0.0479 &  0.0157  &   0.0477& 0.0157\\
                                            &    & 40-50  & 0.1106&  0.0107 &  0.1093 &  0.0104  \\
                                             &    & 60-70  & 0.1827  & 0.0322   & 0.1765  & 0.0279\\
\hline
\multirow{3}{*}{AuAu}  & \multirow{3}{*}{130} & 0-6 & 0.0209  & 0.0095 &   0.0209&  0.0095\\
                                            &    & 34-45&  0.0945 &   0.0246&    0.0936   &0.0241\\
                                             &    & 45-58   &0.1260 & 0.0288   & 0.1239 &0.0284\\
\hline
\multirow{3}{*}{CuCu}  & \multirow{3}{*}{200} & 10-20 &  0.0799&   0.0245&   0.0794&  0.0241\\
                                            &    & 40-50& 0.1796 &  0.0219& 0.1738   &0.0192\\
                                             &    & 60-70 &   0.2580 &  0.0333   & 0.2410  & 0.0238\\
\hline                                             
                                             
\end{tabular}
\end{table*}

The predicted correlation evolves systematically with centrality. In central collisions ($0\text{--}6\%$, $10\text{--}20\%$), the distribution is narrow and tightly clustered near $\alpha_{part} \approx 0$, $y_0 \approx 0$, reflecting the symmetric participant geometry of near head-on events. Moving toward mid-central and peripheral bins ($34\text{--}58\%$, $60\text{--}70\%$), the correlation broadens into a near-linear trend spanning the full $[-0.8, 0.8]$ range, consistent with the larger and more asymmetric participant distributions expected in less central collisions. The model reproduces this centrality-dependent broadening accurately across all bins, though residuals grow modestly in the most peripheral intervals (e.g., Cu+Cu $60\text{--}70\%$, Au+Au $45\text{--}58\%$), where reduced particle multiplicities lead to larger statistical fluctuations.

Table ~\ref{tab:mean_rms_params} shows that for centrality classes withheld from training, the RMS of the residuals is well below the RMS of the true values for both $y_0$ and $\alpha_{part}$, with residual-to-true ratios of roughly 10-45\%. The relative error is largest in the most central bins, where the true distributions are narrowest, and smallest in the mid-central bins (about 10-12\%). The same behavior appears for Au+Au at 62.4 and 130 GeV and Cu+Cu at 200 GeV, indicating that the model generalizes across centrality, system, and energy.

\begin{figure}[htbp]
    \centering
    \includegraphics[width=0.5\textwidth]{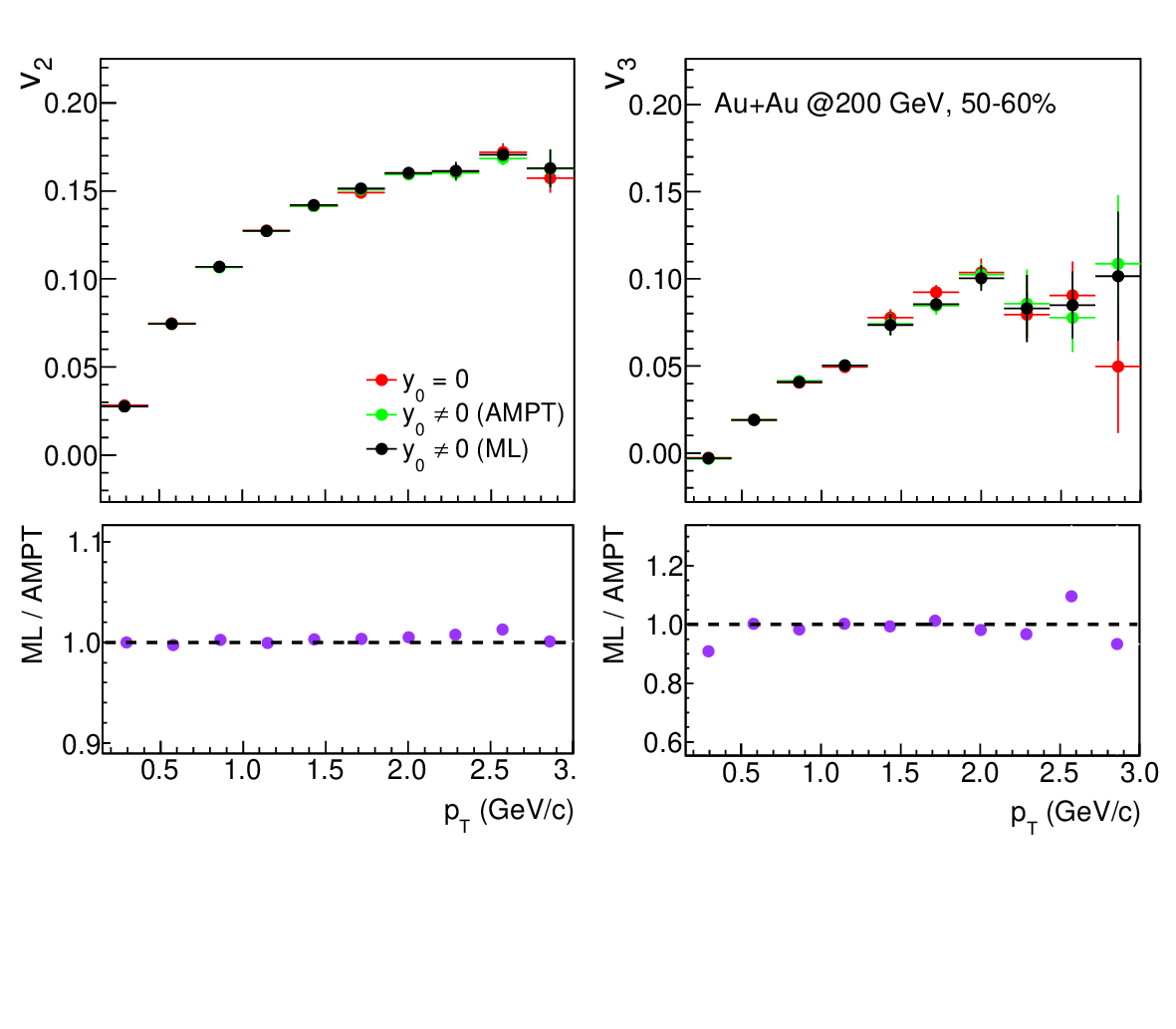}
    \vspace{-1cm}
    \includegraphics[width=0.5\textwidth]{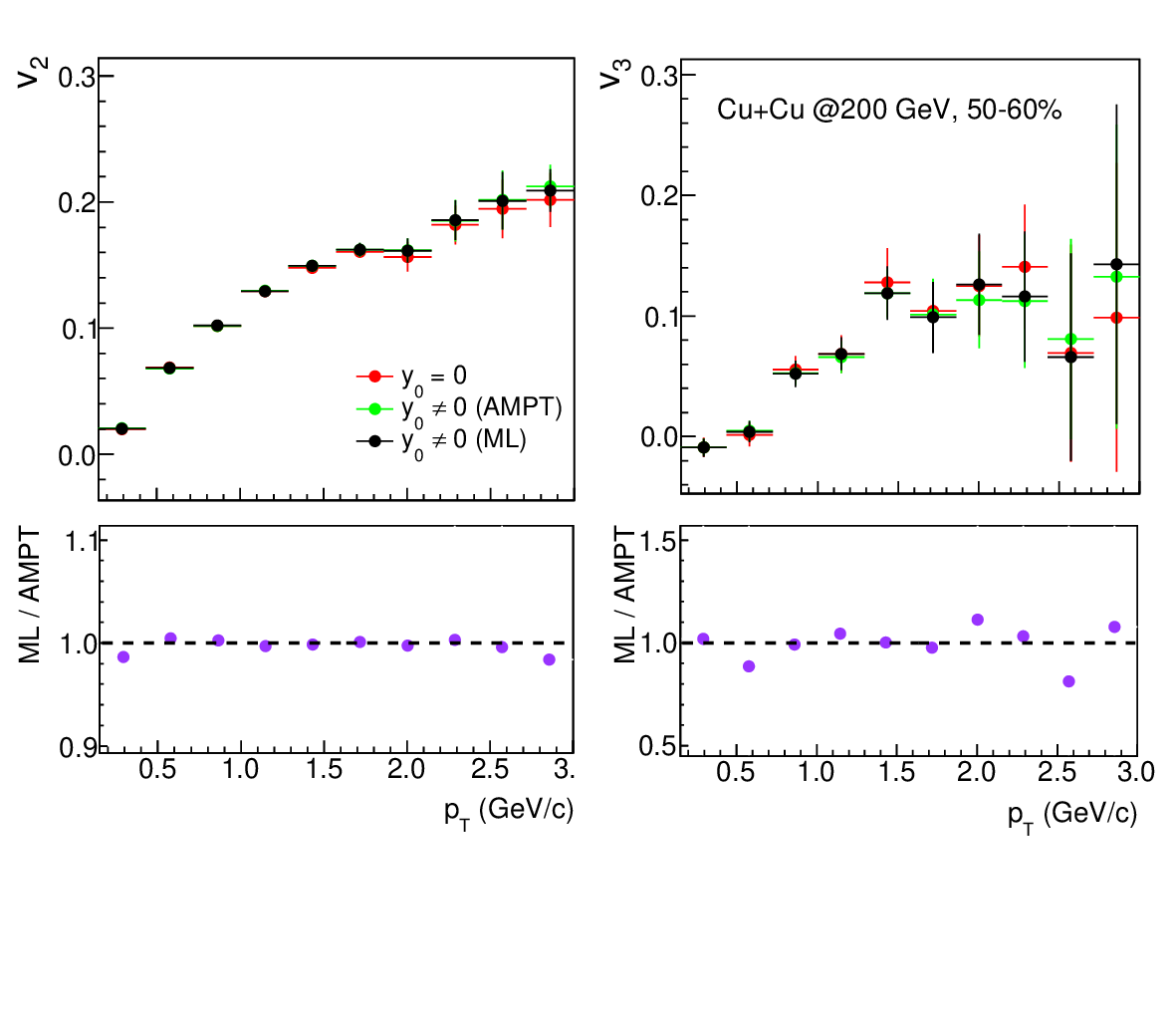}
    \caption{\justifying Effect of the rapidity shift on elliptic and triangular flow in Au+Au and Cu+Cu collisions at $\sqrt{s_{NN}} $  $= 200$ AGeV with comparison between true values and model predictions for 50–60\% centrality class.}
    \label{fig:flow}
\end{figure}

In ultra-relativistic heavy-ion collisions, the spatial anisotropy of the initial-state overlap region is transformed into momentum-space anisotropy through strong,  pressure-driven hydrodynamic expansion. This anisotropic collective motion is traditionally quantified via a Fourier decomposition of the azimuthal particle distribution, where the second and third harmonics represent elliptic flow ($v_2$) and triangular flow ($v_3$), respectively. While elliptic flow primarily reflects the average geometric eccentricity of the nuclear overlap zone, triangular flow originates from event-by-event fluctuations in the initial nucleon positions. The effect of the rapidity shift of the participant zone on the flow observables for Au+Au and Cu+Cu collisions at 200 GeV is illustrated in Figure \ref{fig:flow}. The rapidity shift fluctuates event-by-event, however its effect is not significant in the final flow observables due to the averaging over all the events. The ratio of the flow observables obtained from AMPT and the deep learning model is close to unity, showing reasonably good agreement between these two.
\section{Summary}

We present a fully-connected deep neural network regression model to predict the longitudinal-asymmetry observables $y_0$ and $\alpha_{part}$ from AMPT-simulated heavy-ion collisions. The model is trained on Au+Au and Cu+Cu collision systems at $62.4$ and $200$~GeV, spanning three centrality classes ($0$--$5\%$, $30$--$40\%$, $50$--$60\%$). The architecture of the model is kept intentionally simple: a single fully-connected network with in-graph normalization, trained with a conservative learning rate and MAE loss on event-level-split data. The main methodological emphasis of this work is on the correctness and efficiency on the prediction rather than architectural novelty. 

The model maps six input features through a normalization layer and four hidden dense layers to a two-node linear output. SHAP-based feature attribution identifies collision-system, multiplicity, and $\alpha_{spec}$ as the dominant predictors, while highly correlated but redundant features such as dN/dy contribute negligibly to the output. Prediction resolution is found to degrade systematically towards more peripheral collisions and varies with both beam energy and colliding system, reflecting the reduced per-event particle yield in peripheral and lighter-system collisions. 

True versus predicted comparisons, evaluated separately across all twelve system energy centrality combinations, confirm that the network reproduces both the marginal distributions of $y_0$ and $\alpha_{part}$ and their underlying physical correlation structure with high fidelity, demonstrating the viability of DNN-based regression as a fast, accurate model for extracting longitudinal-asymmetry observables from heavy-ion transport-model output. We also verified the model's generalization capacity by predicting the joint phase-space correlations among $y_0$, $\alpha_{part}$, and $\alpha_{spect}$ across held-out centrality bins ($10\text{--}20\%$, $40\text{--}50\%$, and $60\text{--}70\%$). The model consistently reproduced these multi-variable distributions for Cu+Cu and Au+Au collisions at both $\sqrt{s_{NN}} = 62.4$ and $200\text{ GeV}$ without prior exposure to these centrality ranges during training. The close agreement between AMPT and the DNN for both elliptic and triangular flow further indicates that the ML-based model preserves the kinematic correlations present in heavy-ion collisions.

\section{Acknowledgments}

We acknowledge VECC grid computing facility for the event generation. We specially thank Mr. Prasun Singh Roy and Dr. Vikas Singhal for their support with grid facility.

\end{document}